\documentclass[11pt]{article}
\usepackage{moriond,epsfig}
\usepackage{floatflt}

\def\be{\begin{equation}}
\def\ee{\end{equation}}
\def\bea{\begin{eqnarray}}
\def\eea{\end{eqnarray}}

\def\xgo{x_\gamma^{\rm obs}}
\begin{document}
\vspace*{4cm}
\title{REAL AND VIRTUAL PHOTON STRUCTURE AT HERA}

\author{DORIAN KCIRA}

\address{University of Wisconsin\\ZEUS/DESY, Notkestrasse 85, 22607 Hamburg, Germany\\{\tt dorian.kcira@desy.de}}

\maketitle\abstracts{
The structure of real and virtual photons has been studied in $ep$ collisions at HERA using dijet production.
Measurements of differential dijet cross sections as function of the fraction of photon's momentum
invested in the dijet system are presented. The dependence of the cross sections on the virtuality
of the photon and mean transverse energy squared of the two leading jets has been investigated. QCD calculations
based on current parametrizations of the real and virtual photon parton distribution functions have
been compared to the data.}

\section{Structure of Real Photons}

Interactions of  the real photon ($Q^2 \simeq 0$) have a two-component nature in leading order (LO)
perturbative QCD (pQCD). Thus, two types of hard processes contribute in photon-proton interactions:
in direct photon processes the entire momentum of the photon
takes part in the hard subprocess with a parton from the proton, whereas in resolved photon
processes the photon acts as a source of partons and one of these, carrying a fraction $x_\gamma$
of the photon's momentum, enters the hard subprocess.
Both LO processes can result in the production of two
outgoing partons of large transverse energy that turn into two jets in the final state.
In resolved photon processes, the photon structure is commonly described via parton distribution
functions (PDFs) that receive contributions from both perturbative and non-perturbative
terms. The fraction of photons momentum entering the hard interaction is evaluated using jets
variables~\cite{Ze95}: 

\begin{equation}
\xgo = \frac{E_T^{\rm jet,1}{\rm e}^{-\eta^{\rm jet,1}} + E_T^{\rm jet,2}{\rm e}^{-\eta^{\rm jet,2}} }{2 y E_e}
\end{equation}

\begin{figure}[b]
\epsfig{figure=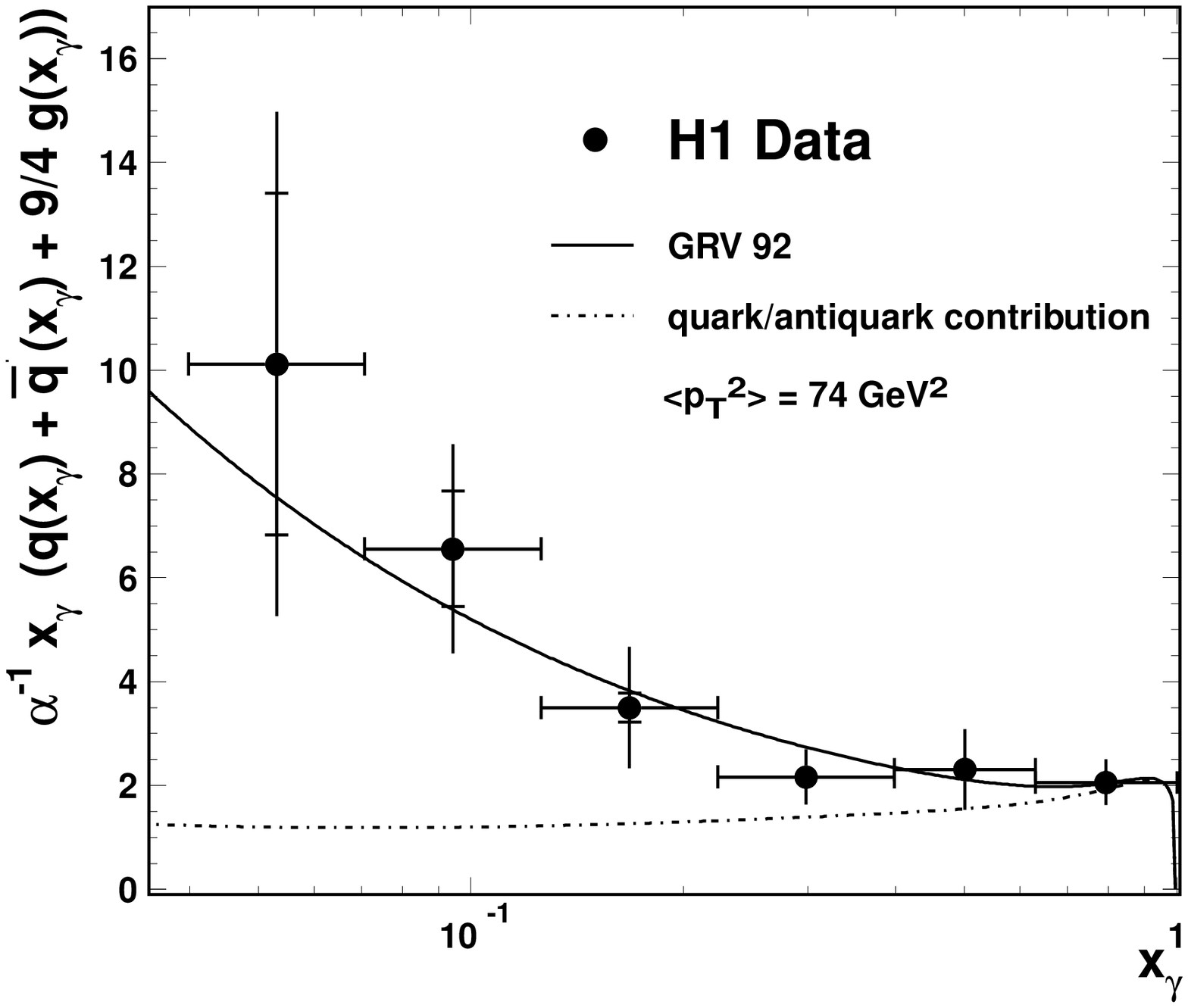,width=8.0cm,height=5.5cm}
\epsfig{figure=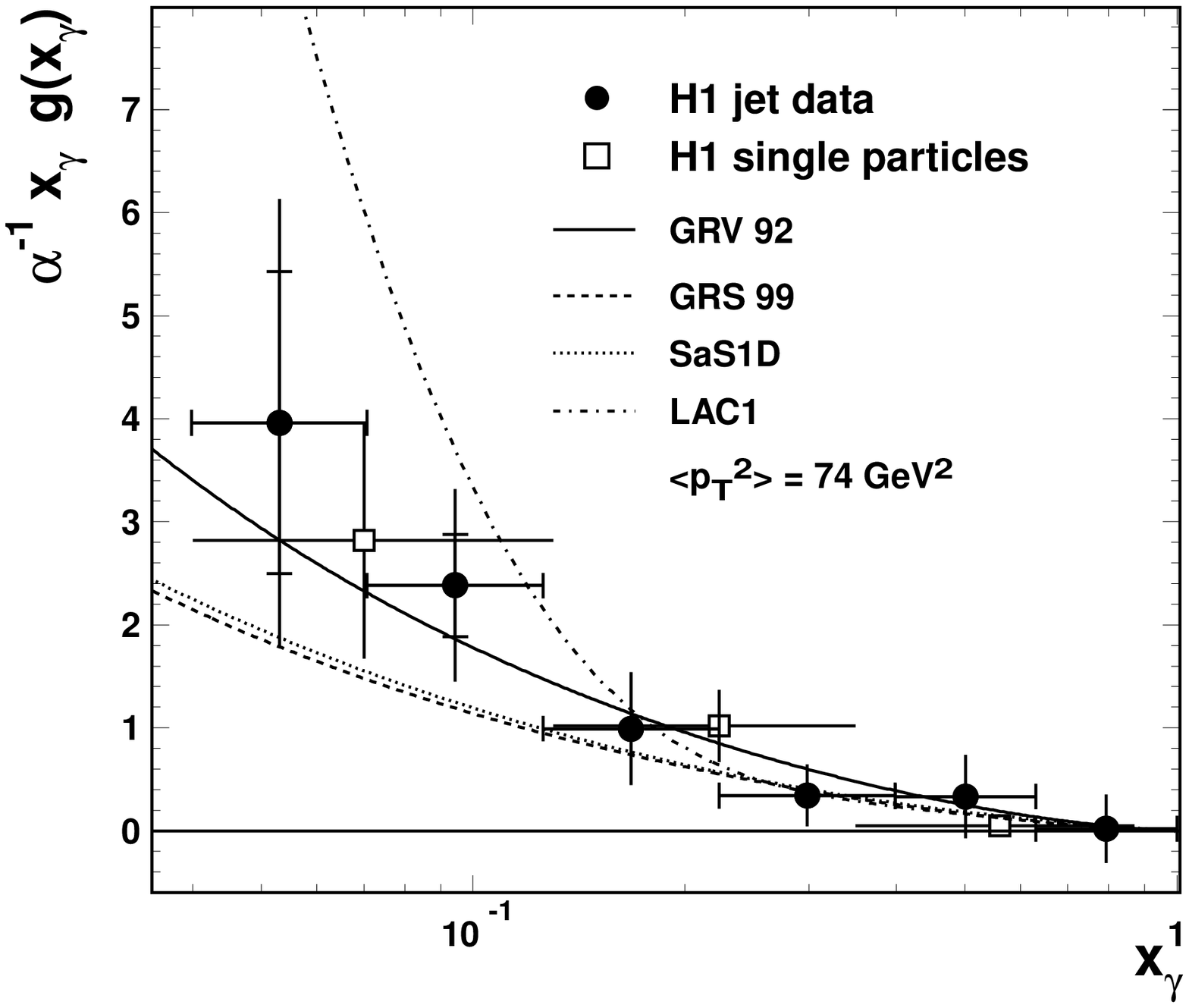,width=8.0cm,height=5.5cm}
\caption{Effective parton distribution (left plot) and gluon distribution (right plot) of the
photon multiplied by $\alpha^{-1}x_\gamma$ as a function of $x_\gamma$. The inner error bars give the
statistical error only and the outer error bars the total error. LO parametrizations of PDFs
based on fits to $\gamma$-$\gamma$ data are also shown.
\label{fig:h1_pdfs}}
\end{figure}

\begin{figure}[t]
\epsfig{figure=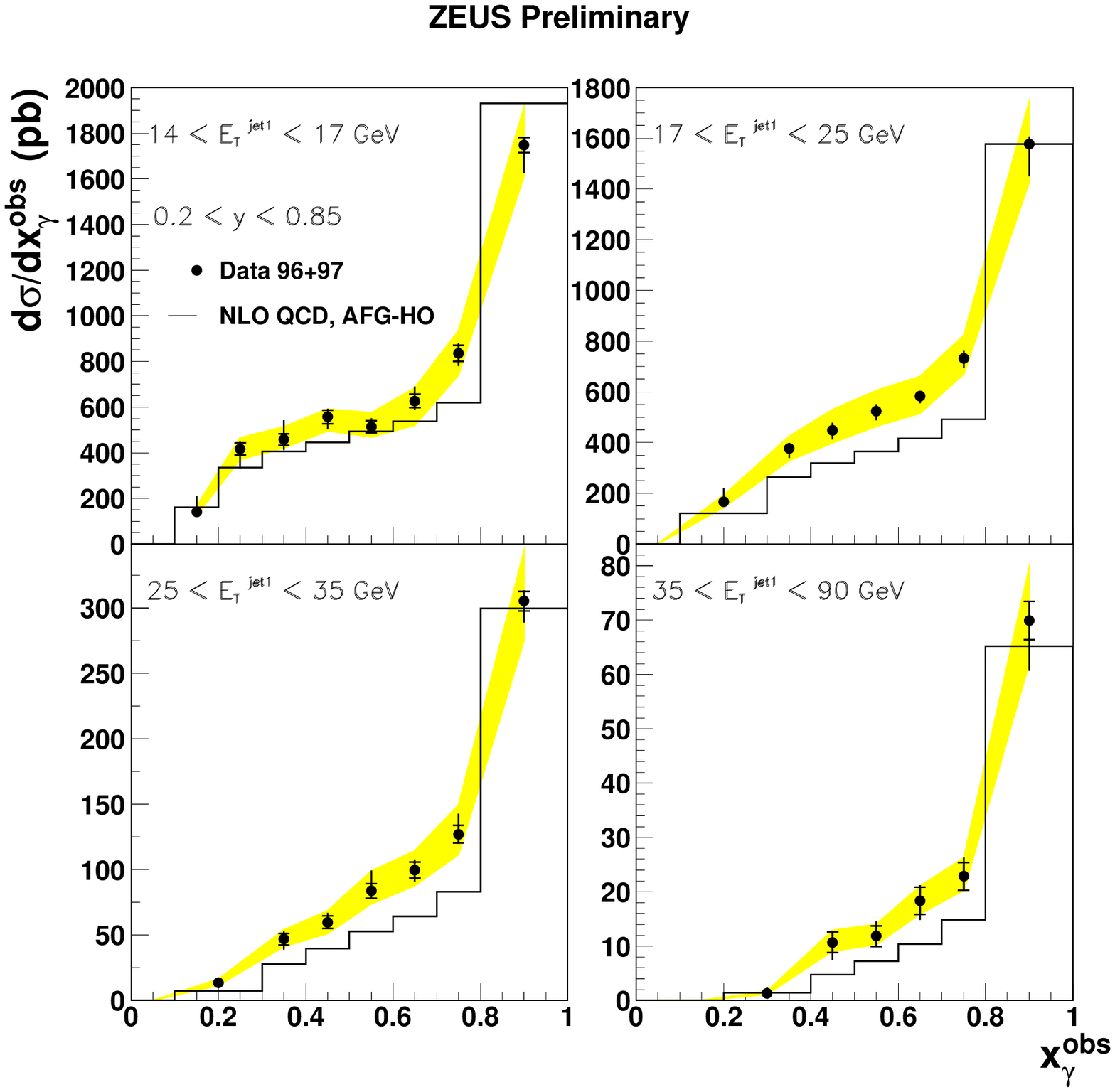,width=8.0cm,height=6.0cm}
\epsfig{figure=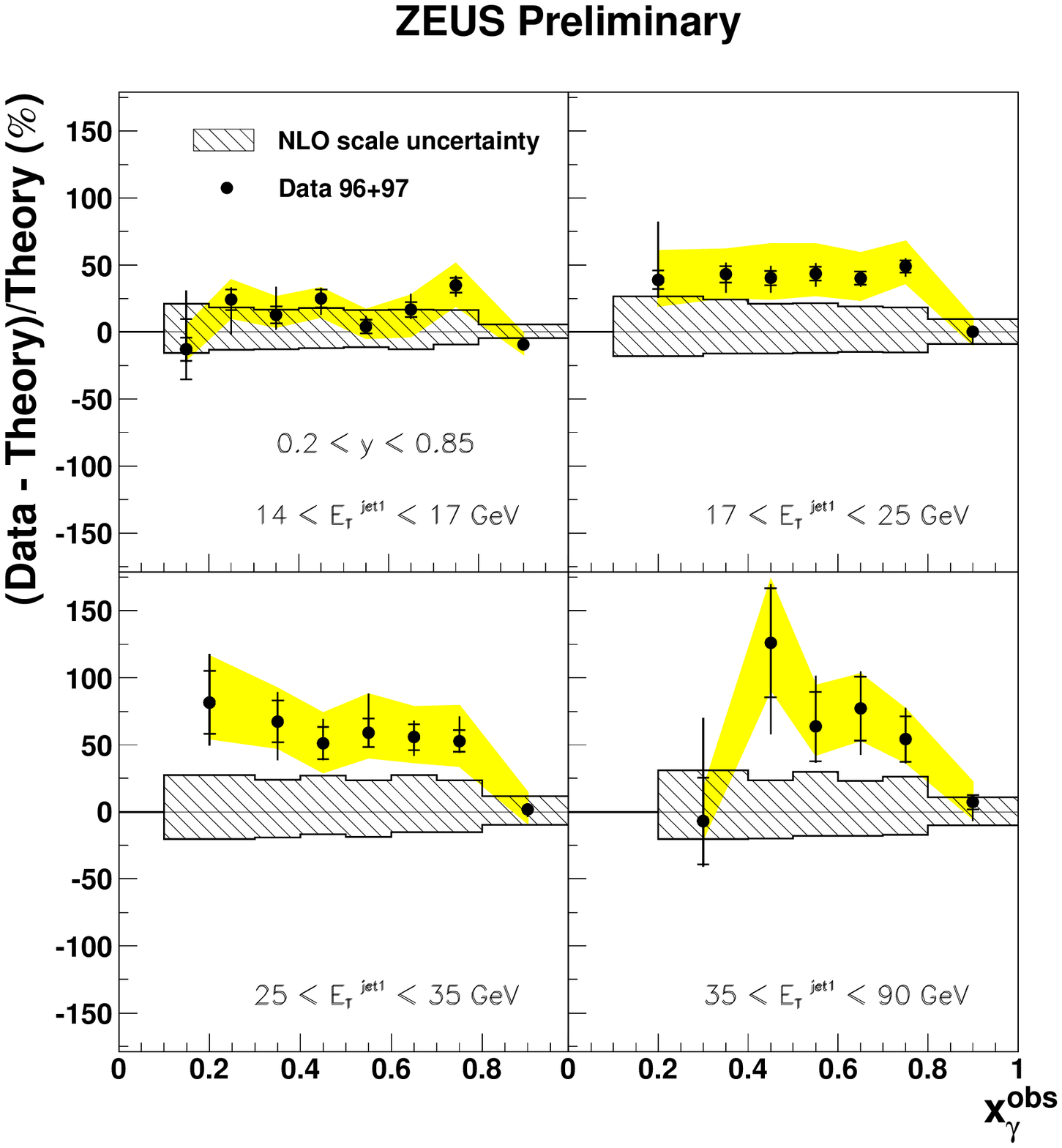,width=8.0cm,height=6.0cm}
\caption{Differential dijet cross section $d\sigma / d\xgo$ in different ranges of the transverse
energy of the jet with the highest $E_T^{\rm jet}$, $E_T^{\rm jet,1}$ (left plot).
The data points are shown with statistical
errors (inner bars) and systematic errors added in quadrature (outer bars). The energy scale
uncertainty is shown as a shaded band. NLO QCD calculations which use
the AFG-HO photon PDFs (histograms) are compared to the data. The percentage differences between
the data and the NLO QCD calculations are also shown (right plot).
\label{fig:zeus_nlo}}
\end{figure}

The H1 Collaboration has determined LO effective parton densities of the photon~\cite{H1r}
from the measurement of the dijet cross section $d \sigma / d \log (\xgo)$. Dijet events were
identified using a cone algorithm with radius $R=0.7$ and selected with jet transverse
energies $E_T^{\rm jet} > 6\;{\rm GeV}$ (after pedestal subtraction) and jet pseudorapidities
$|\eta_{\rm jet,1}-\eta_{\rm jet,2}|<1$, $\eta_{\rm jet}>-0.9-\ln(\xgo)$.
The GRV92 LO pa\-ra\-me\-tri\-za\-tions of the proton and photon PDFs~\cite{Gl92} were used.
An unfolding procedure~\cite{DA95} was used for extracting the effective parton distributions of
the photon: $f_{\rm \gamma,eff} \ = q(x_\gamma) + \overline{q}(x_\gamma) + 9/4 g(x_\gamma)$.
The effective parton density in the photon is adjusted
to get the best agreement with the measured $\xgo$ distribution, having subtracted the LO QCD
expectation for the direct photon contribution, as given by Monte Carlo (MC) simulation.
The measured effective PDF of the photon is shown in figure~\ref{fig:h1_pdfs}.
The contribution of quarks plus antiquarks in the photon as given by the GRV92 parametrization
is shown separately and describes the data
well at the highest values of $x_\gamma$ but falls far below the data at low $x_\gamma$. Within the LO
QCD the difference can only be attributed to a gluon contribution which is shown to
rise strongly towards low $x_\gamma$. The gluon density was then extracted by subtracting
the quark-antiquark contribution as predicted by the GRV92 parametrizations from the extracted
effective parton density (right plot in~\ref{fig:zeus_nlo}).

\begin{figure}
\epsfig{figure=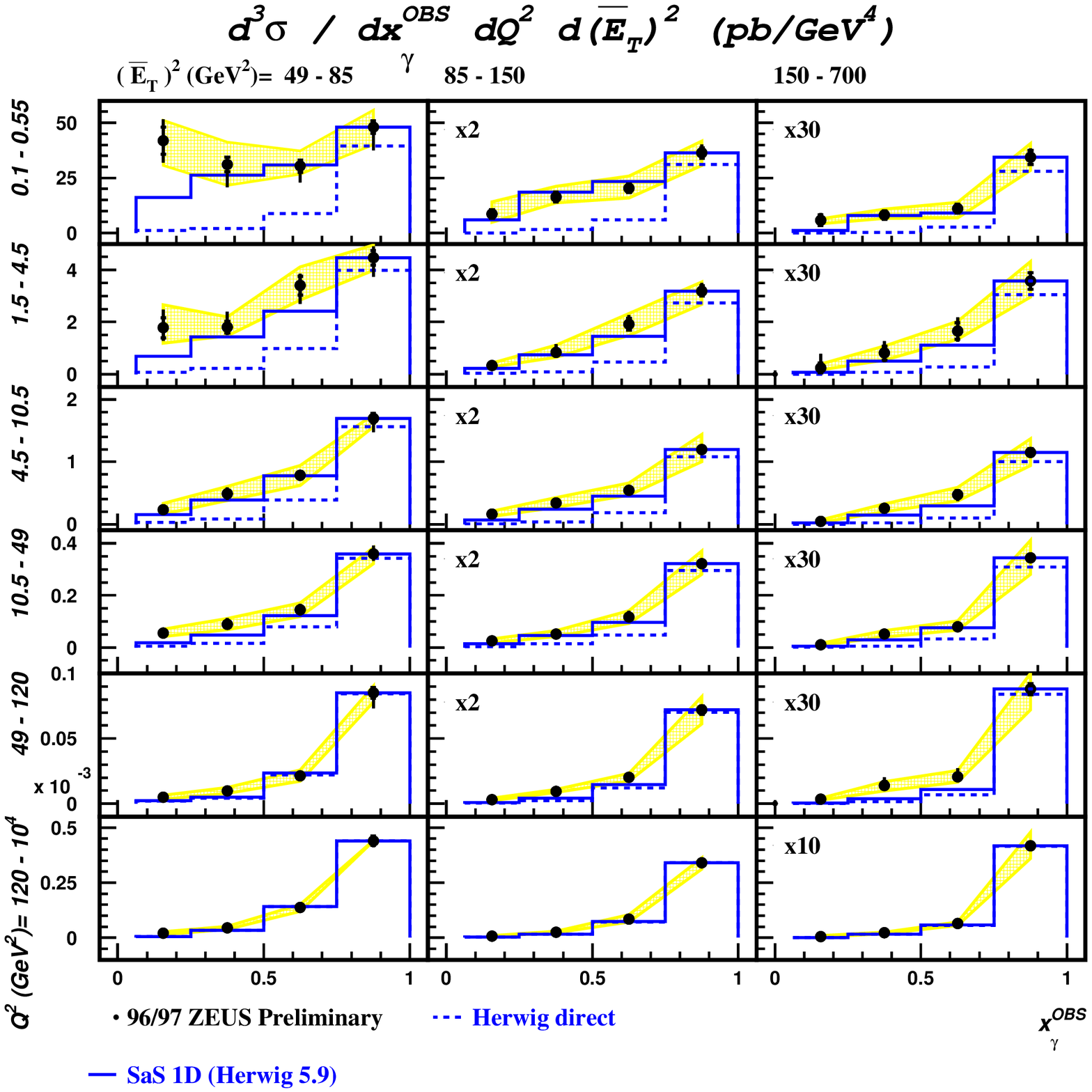,width=8.0cm,height=8.0cm}
\epsfig{figure=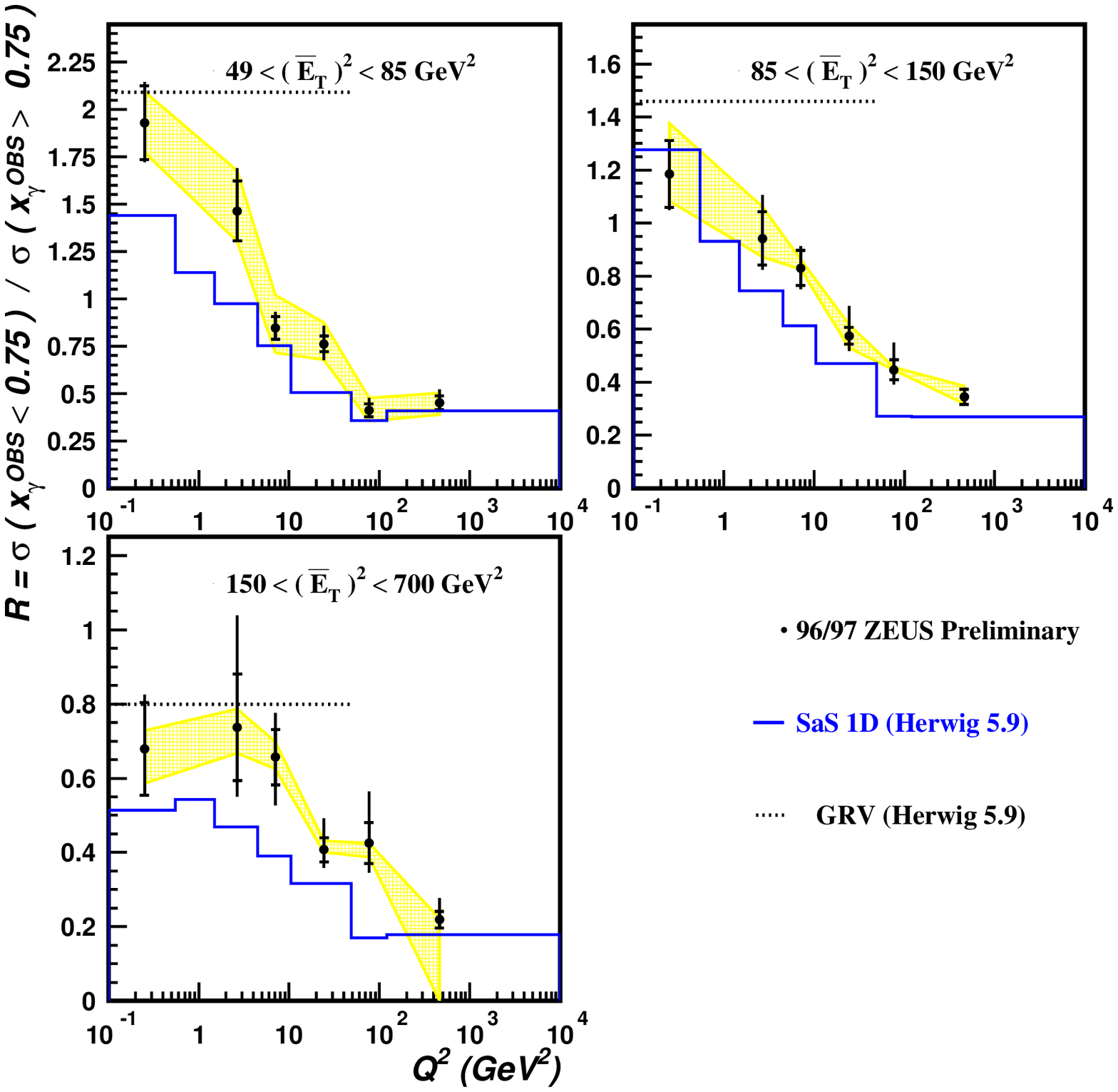,width=8.0cm,height=8.0cm}
\caption{Triple differential cross section $d^3\sigma  / ( d \xgo d Q^2 d \overline{E}_T^2)$
as a function of $\xgo$ for different regions in $Q^2$ and $\overline{E}_T^2$ (left plot).
The ratio of cross sections $R = \sigma(\xgo<0.75)/\sigma(\xgo>0.75)$ as a function of $Q^2$ for
different ranges in $\overline{E}_T^2$ (right plot). The points represent the measured cross
sections with statistical errors (inner error bars) and uncorrelated systematic errors added in quadrature to them
(outer error bars). The shaded bands display the uncertainty in the plotted quantities due to that in
the jet energy scale.
\label{fig:virtual_g}}
\end{figure}

The ZEUS Collaboration has measured dijet differential cross sections $d \sigma / d \xgo$
and compared them to NLO QCD calculations.
Dijet events were identified with the $k_T$-cluster algorithm and selected with $E_T^{\rm jet,1}>14\;{\rm GeV}$,
$E_T^{\rm jet,2}>11\;{\rm GeV}$ and $ -1 < \eta^{\rm jet} < 2$. The cross section was measured in
the kinematic region $Q^2<1\;{\rm GeV}^2$ and inelasticity $0.20 < y < 0.85$.
The measured  differential cross section is shown in figure~\ref{fig:zeus_nlo} in four
ranges of $E_T^{\rm jet,1}$. The data are compared to NLO QCD calculations~\cite{Fi97}
that use the AFG-HO~\cite{Au94} (CTEQ4M~\cite{La97}) parametrization for the PDFs of the photon (proton).
The
data are presented at the hadron level and compared to the calculations at the parton
level. The effect of hadronisation has been evaluated to be $10 - 15 \%$ using the Herwig~\cite{herwig6.2}
and Pythia~\cite{pythia_new} MC programs.

For all regions of $E_T^{\rm jet,1}$, the data at high $\xgo$ are reasonably well described
by the calculation. However, for $\xgo<0.8$ the data are always above the NLO prediction,
most significantly in the higher $E_T^{\rm jet,1}$ regions.
The percentage difference between data and NLO calculation is also shown in
figure~\ref{fig:zeus_nlo}.
The data lie $50 - 60\%$ above the
NLO calculation for $E_T^{\rm jet} > 17\;{GeV}$; this discrepancy is larger than the uncertainties of the measurement
and the estimation of the scale uncertainty in the NLO calculation.
The data therefore suggest inadequencies in the current parametrization used for describing the structure
of the photon. The use of the data in future fits would greatly improve our understanding
of the photon PDFs.

\section{Structure of Virtual Photons}

As the photon virtuality becomes non-zero, the non-perturbative contributions to the
photon PDFs are expected to diminish. In this region, the language of photon PDFs
may still be retained in describing processes where the virtual photon splits into a $q\bar{q}$
pair before interacting with the proton. However, the virtual photon PDFs are in principle
calculable in pQCD. QCD predicts that for $(\overline{E_T})^2 > Q^2$, the virtual photon
PDFs should decrease logarithmically as $Q^2$ grows~\footnote{
$\overline{E}_T$ is the mean transverse energy of the two leading $E_T$ jets.
}.

Dijet events are selected in the $\gamma^* p$ frame with $E_T^{\rm jet,1}>7.5\;{\rm GeV}$,
$E_T^{\rm jet,2}>6.5\;{\rm GeV}$ and $ -1 < \eta^{\rm jet} < 2$.
The cross sections are measured in the phase space defined by
$0.2 < y < 0.55$ and
$0.1<Q^2<0.55$, $1.5<Q^2<4.5$, $4.5<Q^2<10.5$, $10.5<Q^2<49.0$ and $49.0<Q^2<10^4\;{\rm GeV^2}$.

The measured triple differential dijet cross sections $d^3\sigma  / ( d \xgo d Q^2 d \overline{E}_T^2)$
are shown as a function of $\xgo$ in figure~\ref{fig:virtual_g} in different bins of $Q^2$ and $\overline{E}_T^2$.
The LO Herwig predictions using the SaS1D parametrization~\cite{Sc96}
for the photon PDFs do not describe the absolute cross section of the
data. They are therefore normalized to the highest $\xgo$ bin ($\xgo>0.75$) in order to compare
the shape of the data with that of the MC predictions.
For each $\overline{E}_T^2$ bin, the cross section in the low $\xgo$ region falls faster with increasing
$Q^2$ than the cross section in the high $\xgo$ region.
For the bins with $Q^2>\overline{E}_T^2$ the data are well described by the Herwig predictions
including only the LO-direct component. In the bins with $Q^2<\overline{E}_T^2$ the LO-direct
component is not enough to describe the data.

The ratio of cross sections $R = \sigma(\xgo<0.75)/\sigma(\xgo>0.75)$ as a function of $Q^2$  and for
three ranges in $\overline{E}_T^2$ is shown in figure~\ref{fig:virtual_g}.
The ratio of the data falls with increasing $Q^2$ for each range of $\overline{E}_T^2$.
These results show the suppression of the virtual photon structure with increasing photon
virtuality.
The Herwig prediction using SaS1D also falls with increasing $Q^2$ but underestimates the measured
ratio.

\section{Conclusions}

The structure of real and virtual photons has been studied at HERA using dijet production.
The use of the data from the real photon measurements
in future fits would greatly improve our understanding of the photon PDFs.
More thorough studies can be done using the new measurements in the challenging phase space region
of virtual photons.

\end{document}